# Using graph concepts to understand the organization of complex systems

Claire Christensen and Réka Albert

Department of Physics, Pennsylvania State University, University Park, PA 16802, USA

(e-mail cpc146@phys.psu.edu, rza1@psu.edu)

## Abstract

Complex networks are universal, arising in fields as disparate as sociology, physics, and biology. In the past decade, extensive research into the properties and behaviors of complex systems has uncovered surprising commonalities among the topologies of different systems. Attempts to explain these similarities have led to the ongoing development and refinement of network models and graph-theoretical analysis techniques with which to characterize and understand complexity. In this tutorial, we demonstrate through illustrative examples, how network measures and models have contributed to the elucidation of the organization of complex systems.

## Introduction

A wide variety of systems share the feature that they exhibit complex, unpredictable, and perhaps chaotic behavior at the global level, even though they are composed from constituents whose individual interactions are relatively simple and predictable. Such complex systems usually consist of a large number of components and a variety of interactions that can span spatiotemporal scales. The World Wide Web, the Internet, neural networks, social networks, urban street systems, and cellular networks, for example, all constitute complex systems (for reviews, see [Albert and Barabasi, 2002; Albert, 2005; Boccaletti, Latora, Moreno, Chavez and Hwang, 2006]). In the past decade we have witnessed dramatic advances in the understanding of complex systems, prompted by several parallel developments. The computerization of data acquisition in all fields has led to the emergence of large databases on the topology of various real networks, and the slow but noticeable breakdown of boundaries between disciplines has permitted the research and comparison of these diverse networks, bringing to light their shared properties. In addition, increased computing power has allowed us to investigate networks containing millions of nodes—to explore questions that could not be addressed before. This exploration has underscored the need to move beyond reductionist approaches, trying, instead, to understand the behavior of a system as a whole. In cellular biology, for example, the combination of increased computing power and improved experimental techniques has allowed researchers to probe the inner-workings of the cell at an integrated, genome-wide level. The cell serves as a vivid example of a system that exhibits a variety of complex behaviors and properties as a result of the interrelationships among its many components (for a review, see [Albert, 2005]): proteins, genes, metabolites, enzymes, and other entities work together to carry out the basic processes of life. Similar advances have been made in the social sciences, where individuals in a school or workplace are linked to one another through their social relationships [Wasserman and Faust, 1994]; thus disease and rumor spreading can be studied as dynamical processes taking place on these networks [Moore and Newman, 2000; Pastor-Satorras and Vespignani, 2001]. In civil engineering, the complex layout of towns and cities has been greatly simplified by describing urban infrastructure in terms of the web of streets and intersections that comprise a city's skeleton [Scellato, Cardillo, Latora and Porta, 2006]. While the concept of networks and networking has pervaded the social vernacular for some time, and while it is conceptually straightforward to think of many systems as networks, the dedicated mathematical study and analysis of complex networks is an emerging field (see, for example, http://darwin.nap.edu/books/0309100267/html/). In most cases, the question is not how to envision a system as a network, but rather how such a network can be quantitatively described. Network research focuses on three major lines of enquiry:

(i) What are the best measures that capture the most salient features of a network?
(ii) What are the constraints or processes that determine how networks grow and change?
(iii) How does the topology of complex systems influence their dynamics?

In this tutorial we will focus on the first two questions, while the third will be described in detail in two of the following tutorials.

The analytic tool box available for the study of complex systems is rooted in a powerful subfield of mathematics, called *graph theory*, which originated in the eighteenth century work of Euler. A system of elements that interact or regulate each other (a network) can be represented by a mathematical object called a graph [Bollobás, 1979, 1985]. A *graph* is a collection of nodes and edges: the interacting components of the system are reduced to a set of *nodes*, and the interactions among the components are represented by *edges* [Bollobás, 1979, 1985]. For example, the biological processes occurring within a cell can be described in terms of at least four different types of networks, and the cell, itself, can therefore be regarded as a network of networks [Ma'ayan, et al., 2005]. At the genomic level, transcription factors regulate the transcription of genes into mRNA's; by collapsing the transcription factors with the genes that encode them, a *gene-regulatory* network of gene-gene interactions can be constructed [Guelzim, Bottani, Bourgine and Képès, 2002; Lee, et al., 2002; Shen-Orr, Milo, Mangan and Alon, 2002; Luscombe, et al., 2004]. Similarly, post-translational interactions of proteins can be described by a *protein interaction* network [McCraith, Holtzman, Moss and Fields, 2000; Gavin, et al., 2002; Ho, et al., 2002; Lee, et al., 2002; Giot, et al., 2003; Spirin and Mirny, 2003; Han, et al., 2004; Yook, Oltvai and Barabasi, 2004].The enzyme-catalyzed biochemical reactions responsible for cellular metabolism can be portrayed as *metabolic* networks [Feinberg, 1980; Jeong, Tombor, Albert, Oltvai and Barabasi, 2000; Wagner and Fell, 2001; Lemke, Heredia, Barcellos, Dos Reis and Mombach, 2004; Tanaka, 2005]; and the progress of biochemical signals from their point of identification by a sensor protein to their arrival at an intracellular target can also be thought of as a type of *signal transduction* network [Guelzim, Bottani, Bourgine and Képès, 2002; Lee, et al., 2002; Luscombe, et al., 2004; Ma'ayan, Blitzer and Iyengar, 2004; Ma'ayan, et al., 2005].

The directionality of edges within a network may also be important to a description of the system: a network's edges can be either directed or nondirected, as is necessitated by the type of interaction the edges represent. The edges of gene-regulatory networks and the Internet, for example, are directed, since they depict relationships for which the source and target of the interaction are known; conversely, protein-protein interaction networks and social networks are typically nondirected, since these relationships tend to be more mutual.

In addition, depending on the type of network under consideration, the nodes and/or edges might have information attached to them. Nodes can be labeled according to the type of entity they represent, or according to their capacity: in a Kirchhoff electric circuit diagram, for example, resistors are labeled with their resistive capacity, and voltage sources are labeled according to the voltage they supply. A specific example of node diversity is found in *bipartite graphs*, whose nodes belong to one of two classes, and whose edges only connect nodes of one class to nodes of the other class. To simplify the analysis of bipartite graphs, one usually constructs their unipartite projections, preserving only one type of node, and connecting all nodes that share the same neighbor in the bipartite graph. For example, collaboration networks are unipartite projections of bipartite graphs whose two types of nodes are individuals (actors, scientists) and collaborative projects (movies, articles).

Quantitative measures of a connection, including distances, costs, and flows, can be described by edge weights: the edges of a metabolic network are often labeled with confidence levels or reaction speeds [Feinberg, 1980; Lemke, Heredia, Barcellos, Dos Reis and Mombach, 2004]; pipes between junctions in a water transportation system are characterized by their flow capacities. Social networks [Krapivsky, Redner and Leyvraz, 2000; Moore and Newman, 2000; Newman, 2001b; Barabási, et al., 2002; Girvan and Newman, 2002], metabolic networks [Feinberg, 1980; Lemke, Heredia, Barcellos, Dos Reis and Mombach, 2004], food webs [McCann, Hastings and Huxel, 1998; Polis, 1998; Barrat, Barthelemey, Pastor-Satorras and Vespignani, 2004], neural networks [Sporns, 2002; Sporns, Tononi and Edelman, 2002; Latora and Marchiori, 2003], and transportation networks [Berlow, 1999; Li and Cai, 2004; Guimera, Mossa, Turtschi and Amaral, 2005], provide excellent examples of weighted networks.

Mid-twentieth-century work on random graphs [Erdos and Rényi, 1959; Erdos and Rényi, 1960, 1961]—i.e. on networks for which the connections among nodes have been randomly chosen—pioneered many of the basic measures and techniques that would later be extended to the analysis of nonrandom networks. By examining real networks from a graph-theoretical standpoint, it has been firmly established that networks with similar functions have similar graph-theoretical properties (reviewed in [Albert and Barabasi, 2002; Dorogovtsev and Mendes, 2002b; Newman, 2003; Albert, 2005; Boccaletti, Latora, Moreno, Chavez and Hwang, 2006]). Thus, if a list of interactions that describe a system can be compiled, synthesized, or inferred, graph-theoretical analysis and modeling of the system provides a powerful means by which to classify the system, while offering insight into how its local topological features and behaviors give rise to emergent mathematical trends at the global level.

# 1. Basic graph-theoretical measures: tools for networks analysis

The following subsections describe graph-theoretical measures that can characterize the topology of a network at multiple scales of complexity. Sections 1.1 through 1.4 deal primarily with global topological measures, while sections 1.5 through 1.7 explore ways in which the subglobal topology of a network can be probed and quantified.

**1.1 Degree and degree distribution**

The *degree* of a node is the number of edges adjacent to that node; if the directionality of interaction is important, a node's total degree can be broken into an *in-degree* and an *out-degree*, quantifying the number of incoming and outgoing edges adjacent to the node. In an edge-weighted graph one can also define a *node strength*, the sum of the weights of the edges adjacent to the node (Fig. 1). The degree of a specific node is a local topological measure, and we usually synthesize this local information into a global description of the network by reporting the degrees of all nodes in the network in terms of a *degree distribution*, $P(k)$, which gives the probability that a randomly-selected node will have degree $k$ (Fig. 1). The degree distribution is obtained by first counting the number of nodes, $N(k)$, with $k=1,2,3,\ldots$ edges, and then dividing this number by the total number of nodes, $N$, in the network (the same procedure can be employed to find in- and out-degree distributions for a given directed network).

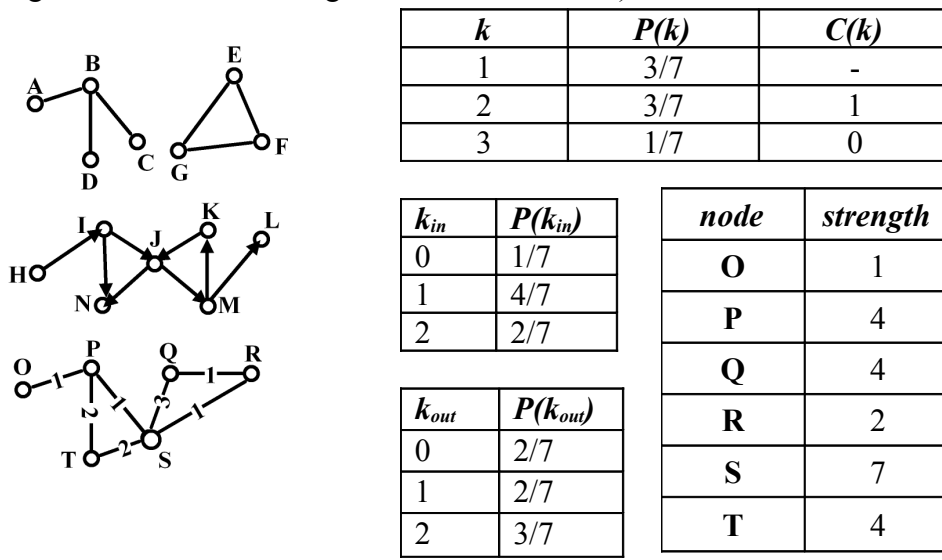

| k | P(k) | C(k) |
|---|------|------|
| 1 | 3/7  | -    |
| 2 | 3/7  | 1    |
| 3 | 1/7  | 0    |

| $k_{in}$ | $P(k_{in})$ |
|----------|-------------|
| 0 | 1/7 |
| 1 | 4/7 |
| 2 | 2/7 |

| $k_{out}$ | $P(k_{out})$ |
|-----------|--------------|
| 0 | 2/7 |
| 1 | 2/7 |
| 2 | 3/7 |

| node | strength |
|------|----------|
| O | 1 |
| P | 4 |
| Q | 4 |
| R | 2 |
| S | 7 |
| T | 4 |

**Figure 1.** The number of interactions in which a component participates is quantified by its (in/out) degree. For example, node *J* has both in-degree and out-degree 2. The clustering coefficient characterizes the cohesiveness of the neighborhood of a node; for example, the clustering coefficient of node *E* is 1, since nodes *G* and *F*, which, together with *E* form a three-node clique, are maximally cohesive. The degree distribution $P(k)$ [$P(k_{in})$ and $P(k_{out})$ in directed networks] quantifies the fraction of nodes with degree $k$, while the clustering-degree function $C(k)$ gives the average clustering coefficient of nodes with degree $k$. If the edges of the network are weighted (bottom graph), the degree is replaced by the strength of the node. The strength of node *P*, for example, is 4. The nodes *J, K,* and *M* of the directed graph constitute a strongly-connected component of the graph. The in-component of the graph contains nodes *H* and *I*, while the out component contains nodes *N* and *L*.

The extensive work of the last decade has demonstrated that networks for similar types of systems have similar degree distributions (reviewed in [Albert and Barabasi, 2002; Dorogovtsev and Mendes, 2002b; Newman, 2003; Albert, 2005; Boccaletti, Latora, Moreno, Chavez and Hwang, 2006]). Although some networks, including the distribution of substations and power lines that forms the North American power grid [Albert, Albert and Nakarado, 2004], and the network of world-wide air transportation routes [Amaral, Scala, Barthelemy and Stanley, 2000] exhibit exponential degree distributions, the vast majority of real networks has been shown to have degree distributions that are *scale-free* (reviewed in [Albert and Barabasi, 2002]) (Fig. 2), meaning that the degree distribution is a power-law:

$$P(k) \sim A k^{-\gamma}. \tag{1}$$

Here, $A$ is a constant that ensures that the $P(k)$ values sum to 1, and the degree exponent $\gamma$ is typically similar for similar networks. Metabolic networks and the out-degree distribution of most gene-regulatory networks, for example, are power-laws with $2 < \gamma < 3$ [Jeong, Tombor, Albert, Oltvai and Barabasi, 2000; Albert and Barabasi, 2002; Guelzim, Bottani, Bourgine and Képès, 2002; Lee, et al., 2002]. The scale-free form of the degree distribution for most real networks indicates that there is a high diversity of node degrees and no typical node in the network that could be used to characterize the rest of the nodes.

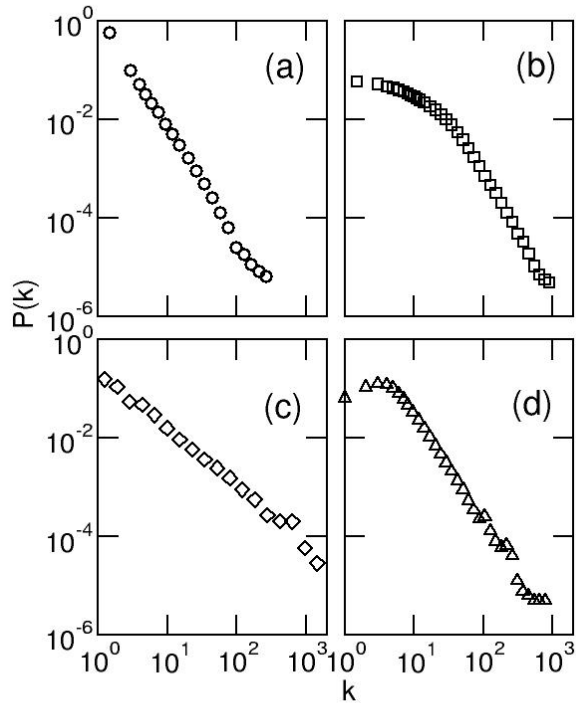

**Figure 2.** The degree distribution of four real networks. *(a)* Internet at the router level. *(b)* Movie actor collaboration network. *(c)* Coauthorship network of high energy physicists. *(d)* Coauthorship network of neuroscientists. Figure reproduced with permission from *Reviews of Modern Physics* (Albert, R. and Barabasi, A.-L., 2002).

is a completely-connected subgraph (*clique*) [Watts and Strogatz, 1998]. Mathematically, the local clustering coefficient is given by

$$C_i = \frac{2E_i}{k_i(k_i - 1)}, \tag{2}$$

where $E_i$ is the number of edges connecting the immediate neighbors of node $i$, and $k_i$ is the degree of node $i$. By averaging the clustering coefficients of all nodes in a network to obtain an *average clustering coefficient*, an idea of the strength of connectivity within the network can be established. Most real networks, including, for example, protein-protein interaction networks and metabolic networks [Wagner and Fell, 2001], as well as collaboration networks in academia and the entertainment industry [Anthonisse, 1971; Granovetter, 1973] exhibit large average clustering coefficients, indicating a high level of redundancy and cohesiveness. Alternatively, the average clustering coefficient of nodes with degree $k$ can be plotted as a function of node degree, $C(k)$ (Fig. 1). It has been found that for a wide variety of networks, this clustering-degree relation has the functional form

$$C(k) = B/k^\beta, \qquad (3)$$

where the exponent β typically falls between 1 and 2 [Ravasz, Somera, Mongru, Oltvai and Barabasi, 2002; Ravasz and Barabasi, 2003; Yook, Oltvai and Barabasi, 2004].

### 1.3 Connectivity, paths, distance, efficiency, and graph components

It is possible that by starting at an edge adjacent to a given node and tracing a *path* along consecutive edges [Bollobás, 1979, 1985], only a fraction of the nodes in the network will be accessible to the starting node. This is often the case in directed graphs, since whether two edges are consecutive depends on their directions. If a path does exist between every pair of nodes in a network, the network is said to be *connected* (if the network is directed, it is said to be *strongly connected*). Often, the *average path length* $d = \langle d_{ij} \rangle$ —i.e. the average number of edges in the shortest path between any two nodes in a network—is used as a characteristic global property of connected networks (Fig. 3a). For most real networks the average path length is seen to scale with the natural logarithm of the number of nodes in the graph: $d \sim ln(N)$. (4)

This *small world* [Watts and Strogatz, 1998] property of real networks implies that path lengths remain small, even if the networks become very large. In the case of directed graphs or unconnected graphs, it is often more advantageous to speak of the graphs' *global efficiency* [Latora and Marchiori, 2001, 2003], or the average of inverse distances,

$$eff = \langle 1/d_{ij} \rangle. \qquad (5)$$

Unconnected nodes' distance is infinite by definition, and thus these node pairs do not contribute to the network's efficiency.

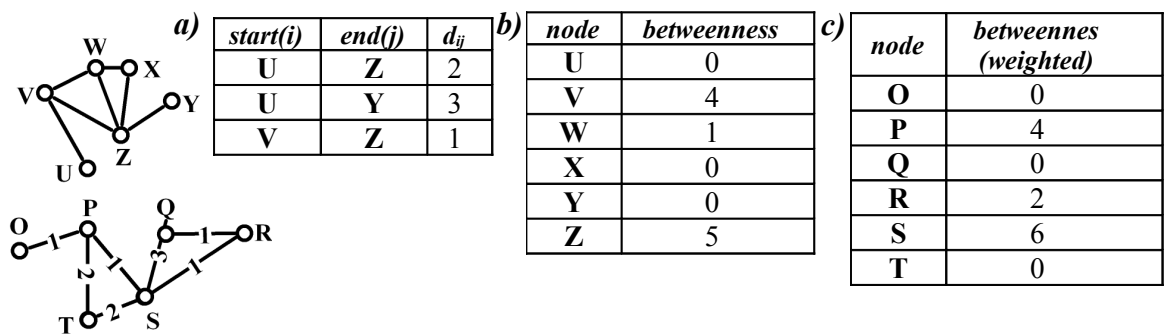

**Figure 3.** **(a)** The graph distance between two nodes is defined as the number of edges in the shortest path between them. For example, the distance between nodes *V* and *Z* is 1, and the distance between nodes *V* and *Y* is 2 (along the *VZY* path). **(b)** The (node) betweenness centrality determines the "traffic" passing through node *i* along all shortest paths between all node pairs in the graph. For example, the betweenness centrality of node *Z* is higher than the betweenness centralities of the other nodes in the graph, since all paths involving node *Y* must pass through node *Z*. **(c)** In a weighted graph, betweenness centrality is calculated by considering least-weight paths, instead of shortest-distance paths. For example, the betweenness centrality of node *S* is 6.

Especially in the case that a network is directed and not strongly connected, it is beneficial to identify connected partitions of the network. For example, a directed network has one or several *strongly-connected components*, a subgraph whose node pairs are connected in both directions. Each strongly-connected component is associated with an *in-component* (nodes that can reach the strongly-connected component, but that cannot be reached from it) and an *out-component* (the converse) (Fig. 1). It has recently been suggested that the nodes of each of these components share a component-specific task within a given network. In signal transduction networks, for example, the nodes of the in-component tend to be involved in ligand-receptor binding; the nodes of the strongly-connected component form a central signaling subnetwork; and the nodes of the out-component are responsible both for the transcription of target genes as well as for phenotypic changes [Ma'ayan, Blitzer and Iyengar, 2004; Ma'ayan, et al., 2005]. By identifying large connectivity classes within a network, one may be able to gain a sense of how the network is functionally organized at an intermediate level of complexity.

**1.4 Betweenness centrality**

While the number and strength of connections associated with a given node is a measure of local centrality, the concept of *betweenness centrality* indicates the node's importance in the overall connectivity of the network [Anthonisse, 1971; Freeman, 1977, 1979]. Node betweenness centrality is defined as the number of (shortest) paths from node $s$ to node $t$ passing through node $i$, divided by the total number of (shortest) $st$-paths (Figs. 3b, 3c); one can similarly define an edge betweenness [Girvan and Newman, 2002]. Both measures of betweenness centrality give some sense of the relative linking and/or traffic-directing capabilities of a node or edge in the graph. As an example, Holme *et al.* have demonstrated that while the most ubiquitous substrates in biochemical pathways do not necessarily have the highest degrees in the network, they often have the highest betweenness [Holme, Huss and Jeong, 2003]. Recently, powerful algorithms (to be discussed in a later section) for determining community structure within a network have been designed with betweenness centrality concepts at their cores. In the following sections, we explore ways in which global measures can be used to deduce information about organizational structures and functions at multiple levels of local complexity.

**1.5 "Extreme" degrees: sources, sinks, and hubs**

As was explained in section 1.1, the degree distribution gives valuable insight into the heterogeneity of node interactivity levels within a network. In directed networks, still more information regarding local organizational schemes can be gleaned by identifying nodes in the network that have only incoming or outgoing edges. These *sinks* and *sources*, respectively, likely have specialized functions. In particular, if the network describes a system in which the flow of information, products, or other entities is of primary focus, the sources and sinks of the network will represent the initial and terminal points of the flow. In signal transduction networks within a cell, for example, extracellular ligands and/or their receptors are typically sources, acting as the injection points for chemical signals; these chemical signals then terminate at effectors-- the networks' sinks [Ma'ayan, Blitzer and Iyengar, 2004]. Vehicular traffic networks may also be characterized by sources and sinks that, in this case, take the form of on- and off-ramps in highway systems [Knospe, Santen, Schadschneider and Schreckenberg, 2002].

In addition, it is also informative to identify the nodes of highest degree, but least abundance, within a network, as their existence greatly constrains the network's topology. These *hubs* have the task of linking multiple paths among nodes: for example, the hubs of a network on which a sexually-transmitted disease propagates are those individuals with the greatest number of sexual contacts [Liljeros, Edling, Amaral, Stanley and Aberg, 2001; Dezso and Barabasi, 2002; Liljeros, Edling and Nunes Amaral, 2003; Grenfell and Bjornstad, 2005]; and the hubs of the World Wide Web are popular webpages to (and from) which multiple other pages are linked [Albert and Barabasi, 2002]. In addition, hubs are generally involved in the largest variety of interactions within the network: the hubs of regulatory networks are often housekeeping regulators and multi-functional transcription factors [Luscombe, et al., 2004]; the dynamics of economic markets are affected by the opinions and actions of socially influential individuals whose interactions in multiple overlapping social networks (e.g. the workplace, housing developments, families, clubs) act as conduits for the transmission of economic attitudes [Erez, Hohnisch and Solomon, 2005]. While the random removal of nodes from the network causes very little disruption (on average) to the connectivity of the network, the targeted removal of the hubs causes the network to quickly break into isolated clusters [Albert, Jeong and Barabasi, 2000]. In the *S. cerevisiae* gene-regulatory network, for example, (individual) knockout of up to 73% of genes has no effect on survival, and thus these genes are not essential to the overall functioning of the organism [Giot, et al., 2003], while the loss of hub genes such as the p53 tumor suppressor gene has dramatic deleterious effects [Vogelstein, Lane and Levine, 2000]**.**

**1.6  From local organization to large-scale clustering:  network motifs**

Network *motifs*—patterns of connection that recur statistically more frequently than they would in a degree-preserving randomized graph [Milo, Itzkovitz, Kashtan, Chklovskii and Alon, 2002; Shen-Orr, Milo, Mangan and Alon, 2002]-- demonstrate that in most real networks, structural and functional organization exists at a level of complexity of only a handful of nodes, and, moreover, that networks with similar overall function appear to be built from the same basic motifs. The high average clustering coefficient of many networks indicates a high abundance of triangles (three-node cliques); in the last five years a significant number of additional interaction motifs have been reported. For example, interaction motifs such as autoregulation and feed-forward loops have a high abundance in transcriptional regulatory networks[Shen-Orr, Milo, Mangan and Alon, 2002; Teichmann and Babu, 2004; Balázsi, Barabási and Oltvai, 2005]; feed-forward loops, feedback loops (both positive and negative) [Ma'ayan, Blitzer and Iyengar, 2004], and triangles of scaffolding (protein) interactions are also abundant in signal transduction networks; feedback loops are common features of the World Wide Web [Barabási and Albert, 1999; Ravasz and Barabasi, 2003]; and a plethora of motifs, including feed-forward and feedback loops, bi-parallels, and bifans have been observed in electronic circuits [Brglez, Bryan and Kozminski, 1989; Cancho, Janssen and Sole, 2001; Ravasz and Barabasi, 2003] (Fig. 4).

| Motif | Motif Name | Network Type |
|---|---|---|
| 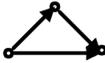 | Feed-forward loop | -Gene-regulatory [1,2,3]<br>-Neural [2]<br>-Electronic circuits [2,4,5,6] |
| 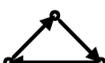 | Feedback loop | -Electronic circuits [2,4,5,6]<br>-Signal transduction [2,7]<br>-World Wide Web [2,6,8] |
| 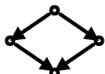 | Bi-parallel | -Neural [2]<br>-Food webs [2,9]<br>-Electronic circuits [2,4,5,6] |
| 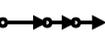 | Chain | -Signal transduction [2,7]<br>-Food webs [2,9] |
| 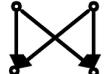 | Bifan | -Gene-regulatory [1,2]<br>-Neural [2]<br>-Electronic circuits [2,4,6] |

**Figure 4.** Common network motifs and network classes (with references) in which they are found. Reference key: 1: Costanzo *et al.*, 2001; 2: Palla *et al.*, 2005; 3: Shen-Orr *et al.*, 2002; 4: Brglez *et al.,* 1989; 5: Cancho *et al.*, 2001; 6: Ravasz & Barabasi, 2003; 7: Ma'ayan *et al.*, 2004; 8: Barabasi & Albert, 1999; 9: Williams & Martinez, 2000.

Many of the information-propagating and processing functions of the motifs identified in networks ranging from gene-regulatory networks to the World Wide Web can be interpreted in terms of the roles these motifs play in electronic circuits [Milo, et al., 2004]. The feed-forward loop, for example, is thought to be involved in persistence detection and pulse generation, and often dictates the acceleration and/or delay of signal responses within a network [Shen-Orr, Milo, Mangan and Alon, 2002; Mangan and Alon, 2003; Ma'ayan, Blitzer and Iyengar, 2004]. The feedback loop, on the other hand, is responsible for state locking [Milo, et al., 2004] and tends to appear in networks in which signaling thresholds are important [Ma'ayan, et al., 2005].

Another point of interest regarding motifs is the fact that these subglobal topological structures should predict and be predicted from the global topology of a given network [Vazquez, et al., 2004]. Motif structure is intimately connected to clustering and degree, and for this reason, it is natural that there should exist a correlation between the degree distribution exponent of a network and the type and abundance of motifs found in that network [Vazquez, et al., 2004]. Alternately, the argument can be made that networks are built from motifs, and that the degree distribution is therefore fundamentally a function of motif abundance. Whichever perspective is adopted, it is straightforward to demonstrate that small motifs tend to cluster around nodes of high degree [Dobrin, Beg and Barabási, 2004]—a feature of real networks that acts as a link between the local network structure and the global topology, and that thereby synthesizes the bottom-up and top-down views of network structure and function. Interestingly, a growing body of evidence suggests that the statistical abundance of motifs in real networks may, in fact, be the result of the dynamic processes involved in network evolution [Valverde and Sole, 2005], and may therefore be only weakly linked to functional constraints.

**1.7 Modularity**

The disconnected clusters that result from the removal of hub nodes within many real networks hint at the possibility of (relatively) densely-connected subglobal node groupings, or *modules* within networks.  The concept of modularity depends less on gross connectivity than it does on density of connection:  that is, modules should not necessarily be understood as disconnected components, but rather as components that have dense intra-component connectivity but sparse inter-component connectivity.  Ultimately, these subgraphs should be functionally-separable, such that different processes within the network will be regulated by distinct modules [Teichmann and Babu, 2004].  Cellular networks, for example, have long been thought to be modular, composed of functionally-separable subnetworks corresponding to specific biological functions [Holme, Huss and Jeong, 2003].  Several methods have been proposed to identify the modules of interaction networks [Girvan and Newman, 2002; Rives and Galitski, 2003; Guimera and Amaral, 2005; Ma'ayan, et al., 2005].  The algorithm of Girvan & Newman is based on the idea that network modules integrate their individual functions through crosstalk along edges of high betweenness centrality [Girvan and Newman, 2002; Knospe, Santen, Schadschneider and Schreckenberg, 2002].  This divisive algorithm breaks the network into modules by identifying the bridges-- the high-betweenness edges-- among the different modules (see Fig. 5). It is also possible to implement agglomerative algorithms, such as the *clique percolation* method recently proposed by Palla *et al.* [2005], to identify modules within a network.  The clique percolation algorithm begins by locating a $k$-clique (completely-connected subgraph on $k$ nodes), since edge density is maximal in a completely-connected graph.  This $k$-clique is then "rolled" through the network to locate other $k$-cliques that share $k$-1 of their nodes with the original $k$-clique.  The group of nodes and edges identified during this process is then regarded as a module [Palla, Derenyi, Farkas and Vicsek, 2005].

Ravasz *et al.* [2002] have recently argued that taken together, a heterogeneous degree distribution, inverse correlation between degree and clustering coefficient (as seen in metabolic and protein interaction networks) and modularity suggest *hierarchical modularity*, in which modules are made up of smaller and more cohesive modules, which themselves are made up of smaller and more cohesive modules, etc.  This type of hierarchical self-similarity may play an important role in network assembly and evolution.  Moreover, identification of hierarchical modules within a network allows for the synthesis of organizational and functional schemes at multiple levels of complexity.  Erikson, *et al.* [2005] have also noted that a comparison of inter- and intra-module links can allow for predictions of mixing and diffusion rates within a network, and can provide an accurate picture of the stability of a large network in the face of edge or node removal.

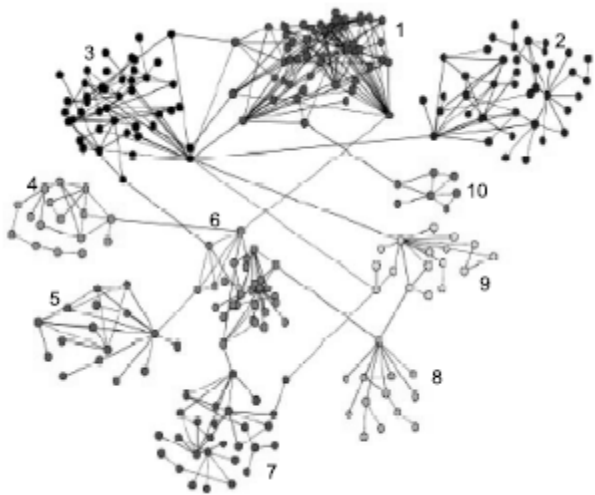

**Figure 5:** The community structure of econophysicist collaboration networks, uncovered by the Girvan-Newman edge-betweenness algorithm. Communities 1,4, and 10 include scientists from Boston University, the University of Southern California and the University of California at Los Angeles, and Jerusalem University, respectively. Community 9 is composed of scientists who are geographically separated, but who work on the same topic (financial markets). Figure reprinted from *Physica A*, doi:10.1016/j.physa.2005.11.018 , P. Zhang, *et al.*, *The analysis and dissimilarity comparison of community structure*, copyright 2005, with permission from Elsevier.

# 2 Network models

Although Erdős-Rényi (ER) random graph theory, the pure mathematics out of which contemporary network analysis and modeling grew, serves as a basis for new graph-theoretical measures and novel network models, random graph theory cannot account for the dominant topological features exhibited by most real-world networks: Nature is decidedly non-random. Thus, most network models have developed out of a need to match theory to observation—to explain the most salient topological features and trends found in real-world networks by modeling the formative mechanisms that explain their existence.

**2.1 Random graphs**

The simplest approach to analyzing complex networks with a seemingly random topology is to determine their expected properties based on the number of nodes and edges alone. For such an analysis, we turn to *random graph theory*, a well-developed branch of discrete mathematics founded on the pioneering work of Pál Erdős and Alfréd Rényi [Erdos and Rényi, 1959; Erdos and Rényi, 1960, 1961]. The theorems of ER random graph theory describe beautifully the qualitative changes in graph topology due to an increase in the number of edges in a graph, but they cannot account for the properties exhibited by real networks. An Erdős- Rényi random graph is formed by randomly connecting $N$ nodes with $E$ edges (Fig. 7a). For large $N$, the degree distribution of such a graph is Poissonian, implying that most nodes have degree $k$, close to the average degree in the graph, $<k>=2E/N$. Therefore, unlike most real networks, for which the degree distribution is scale-free, ER random graphs are largely homogeneous in degree (Fig. 6). In addition, the average clustering coefficient of ER random graphs scales inversely with the size of the network, such that $<C> = <k>/N$, (6)
and the clustering coefficient distribution of an ER random graph is independent of degree, peaking at a value equal to the connection probability $p$. This is not the case for real networks, for which the average clustering coefficient tends to be high and is derived from a distribution that scales inversely with degree. Finally, the average path length of ER random graphs, $\langle d \rangle \approx ln(N)/ln\langle k \rangle$, (7)
remains small, even when the network is large [Bollobás, 1979]; while a variety of real networks have average path lengths close to what is found in random graphs with the same number of nodes and edges, other networks, such as the World Wide Web, have considerably longer average path lengths (reviewed in [Albert and Barabasi, 2002]).

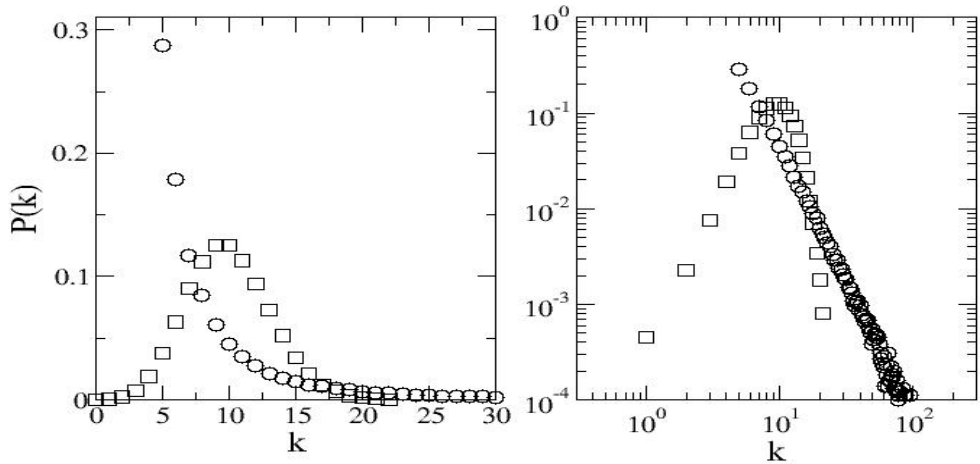

**Figure 6.** Comparison between the degree distribution of scale-free networks (circles) and random graphs (squares) having the same number of nodes and edges. For clarity the same two distributions are plotted both on a linear (left) and logarithmic (right) scale. The bell shaped degree distribution of random graphs peaks at the average degree and decreases quickly for both smaller and larger degrees, indicating that these graphs are statistically homogeneous. In contrast, the degree distribution of the scale-free network follows the power law $P(k) = Ak^{-3}$, which appears as a straight line on a logarithmic plot. The continuously decreasing degree distribution indicates that low-degree nodes have the highest frequencies; however there is a broad degree range with non-zero abundance of very highly connected nodes (hubs) as well. Note that the nodes in a scale-free network do not fall into two separable classes corresponding to low-degree nodes and hubs, but every degree between these two limits appears with a frequency given by *P(k)*. Figure reproduced with permission from *Journal of Cell Science* (Albert, R., 2005).

A subsequent family of graph-formative models replaced the Poissonian degree distribution of the ER model with a prescribed degree distribution, according to which nodes are randomly attached to one another [Newman, 2001a]. These *scale-free random graphs* exhibit similar qualitative topological transitions as ER random graphs do [Newman, 2001a], and their average clustering coefficient and path length are similar to those of random graphs [Newman and Watts, 1999]. Scale-free random graphs with an underlying bipartite structure can at least partially explain the high clustering coefficients of collaboration networks, and they serve as effective null models when searching for non-random network features at the sub-global level [Newman, 2001b]. However, these graphs make no attempt to account for why real networks are scale-free.

**2.2 The Watts-Strogatz model**

The first model to capture the short path lengths, and large, graph-size-independent average clustering coefficients of real networks was proposed by Watts and Strogatz [1998]. Their model interpolates between an ordered, finite-dimensional lattice, whose clustering coefficient is independent of the size of the lattice, but whose path lengths are long, and a random graph, whose clustering coefficient is not size-independent, but whose path lengths are closer to real-world values. The Watts-Strogatz (WS) model postulates that because real networks seem to have topological features that lie somewhere between those of a lattice and those of a random graph, by starting with a lattice and rewiring the edges with probability $p$ (avoiding self-connections and repeated edges), a network with the size-independent clustering coefficients of lattices and the small average path lengths of random graphs will evolve [Watts and Strogatz, 1998](Fig. 7b). Because, due to rewiring, it is possible that portions of a WS model network will become disconnected, Newman and Watts also suggested a variant model where instead of rewiring, edges between randomly selected lattice sites are added with a probability $p$; this variant has similar properties as the original for small $p$[Newman and Watts, 1999]. Although the WS model does produce regions for which there is both a large clustering coefficient and small average path lengths, the model is difficult to compare to real networks, because the rewiring probability, $p$, is generally not known (or is nonexistent) in real networks. In addition, the degree distribution of the model network is centered about the average degree for any rewiring probability, thus causing the degree distribution to more closely resemble the Poissonian degree distribution of random graphs than the scale-free degree distribution of real networks.

**2.3 Evolving network models**

Neither random graph theory nor the Watts-Strogatz model can reproduce the scale-free degree distribution of most real networks. Accounting for this feature necessitates a shift from modeling network topology to modeling network assembly. Since real networks are often the product of some type of developmental process, be it, for example, biological evolution in the case of cellular networks, or economic evolution in the case of trade networks, capturing the mechanisms behind a network's assembly has the potential of elucidating both its current and future topology. The first successful model that could account for the natural evolution of a scale-free degree distribution—the model that is still at the heart of many evolving graph models today—was the Barabási-Albert (BA) model [Barabási and Albert, 1999], which has at its core two essential elements: growth and preferential attachment. The BA model demonstrates that by adding nodes to a network—i.e. by allowing the network to grow—and by assigning new edges according to a "rich-get-richer" scheme, such that nodes with high degrees have a greater chance of acquiring still more links, a network with a power-law degree distribution will develop naturally (Fig. 7c). While the BA model marks a significant step in the theory of graph evolution, it does not accurately describe the topology of real networks, as the degree-distribution exponents of BA graphs are on the high end of what is observed in real networks, their clustering coefficients and path lengths are smaller than those of real networks [Bollobás and Riordan, 2003], and they have constant clustering coefficient distributions [Ravasz and Barabasi, 2003].

The success of the BA model in producing a scale-free degree distribution by dynamics, as opposed to by design, lead to a frenzy of activity in developing evolving graph models that would not only capture the scale-free nature of real networks, but that would also come closer to predicting more realistic clustering coefficients and distances. Most of these models were still based on the ideas of growth and preferential attachment, but included the addition of features such as nonlinear attachment [Krapivsky, Redner and Leyvraz, 2000], initial attractiveness of isolated nodes [Jeong, Neda and Barabasi, 2001], accelerated growth [Dorogovtsev and Mendes, 2002a], aging [Dorogovtsev and Mendes, 2002b], and fitness [Bianconi and Barabási, 2001b, a]. These variations on the BA model have seen significant success in producing networks with realistic properties, and led to a better understanding of the transitions between scale-free, single-scale (exponential) and hub-and-spoke network topologies.

A recent model, proposed by Ravasz *et al.*, producing networks with power-law degree and clustering coefficient distributions, is based on a self-similar growth pattern and not on preferential attachment [Ravasz, Somera, Mongru, Oltvai and Barabasi, 2002]. This model starts with a densely-connected seed graph on a small number (*n*) of nodes. The seed graph is then multiply replicated, and the outer nodes of the resulting construction are connected to a central node (Fig. 7d). This process is repeated until the desired graph size is reached. The net result of the replication model is a degree-distribution exponent

$$\gamma = 1 + \frac{log(n)}{log(n-1)} \qquad (8)$$

. For small *n*, the replication model produces a degree-distribution exponent very close to what is seen in, for example, metabolic networks, and, further, the clustering coefficient is seen to scale inversely with node degree. However, as successful as this model is, it permits only certain degrees and clustering coefficients, while in most real systems, gaps in degree and/or clustering coefficient values are usually not observed.

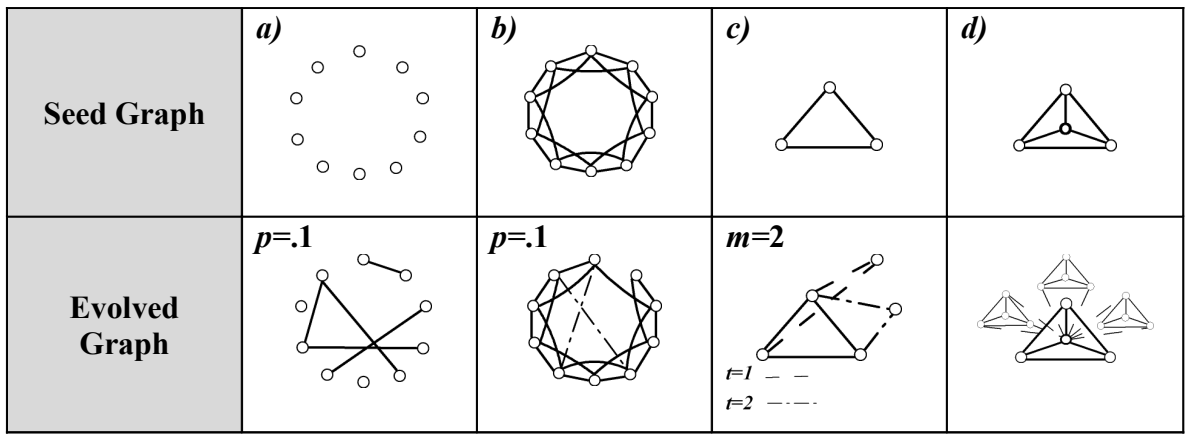

**Figure 7:** Graph evolution models. **(a)** Erdös-Rényi random graph model: with probability *p*, nodes are linked at each time step. **(b)** Watts-Strogatz model: at each time step, edges are randomly rewired with probability *p*. **(c)** Barabási-Albert model: at each time step, a node and *m* edges are added to the graph, such that the new edges are most likely to terminate on nodes of already high degree. **(d)** Ravasz *et al.* replication model: at each time step, the graph is replicated, and the outer nodes of the replicant are attached to the central node of the seed.

An emerging body of research has demonstrated that complex networks with (predominantly) scale-free degree distributions, high average clustering coefficients, and small-world distances may result from a melding of dynamics and optimization on a network of fixed size [Cancho, Janssen and Sole, 2001; Colizza, Banavar, Maritan and Rinaldo, 2004; Valente, Sarkar and Stone, 2004]. Graph optimization models have largely focused on creating networks for which some combination of the average degree and the average distance between nodes in the network is minimized [Cancho, Janssen and Sole, 2001; Valente, Sarkar and Stone, 2004]. Recent schemes have also sought to maximize network resilience (both to random removal of nodes, as well as to targeted attack of the hubs) by generating degree distributions that ensure the maintenance of a giant connected component (a cluster whose size scales linearly with the size of the network). Most optimization schemes begin with an undirected graph with a fixed number of nodes and/or links. An optimization function is defined, and then the edges of the network are randomly rewired with some probability, $p$. If, post-rewiring, the value of the optimization function is more optimal than its previous value, the rewired state of the network is kept, and the algorithm is rerun until no further optimization is possible. This type of optimization procedure has been very successful in generating networks with real-world qualities that can account for such natural phenomena as the formation of channel networks in river basins and the emergence of allometric scaling in biological systems [Banavar, Maritan and Rinaldo, 1999; Colizza, Banavar, Maritan and Rinaldo, 2004].

Recently, other work has modeled network growth as responding to and being constrained by the physical space in which the network exists (reviewed in [Hayaski, 2006]). The nodes of these *spatial networks* are embedded in Euclidean space, and the edges among the nodes are representative of real distances. This type of modeling has proven successful for networks whose topologies are strongly correlated to the physical space in which the networks exist; for example neural cells[Sporns, 2002; Sporns, Tononi and Edelman, 2002], the router and autonomous system level topology of the Internet [Yook, Jeong and Barabasi, 2002], and urban infrastructure maps [Marchiori and Latora, 2000; Crucitti, Latora and Porta, 2005]. Spatial embedding imposes a maximal degree on any one node in the network, and the probability of connection decreases with increasing distance between nodes; for this reason spatially-embedded graphs are not small worlds [Csanyi and Szendroi, 2004] and they exhibit a clustering coefficient distribution that is independent of degree [Ravasz and Barabasi, 2003]. These properties suggest that spatially-embedded networks represent a new class of graph objects that may have evolutionary mechanisms different from those of nonspatial scale-free networks.

# Network topology and dynamics

In the past decade, through empirical studies, modeling, and analysis, we have learned that markedly different complex systems display generic organizing principles. Yet, we believe these results are only the tip of the iceberg. As more data becomes available, scientists will continue to refine graph-theoretical measures and augment the classification of real networks. In addition, recent modeling of dynamic processes on networks [Albert and Othmer, 2003; Glass, Kappey and Grenfell, 2004; Sood and Redner, 2005] promises a rich future for the study of the interplay between networks, nonlinear dynamics, and chaos. There is increasing evidence that the dynamical behavior of complex systems is profoundly influenced by their topological organization. For example, the topology of human contact networks, the substrate of disease spreading, determines the emergence of epidemics. Pastor-Satorras and Vespignani [2001] showed that while for random networks a local infection spreads to the whole network only if the spreading rate exceeds a critical value, for scale-free networks any spreading rate leads to the infection of the entire network.  This finding has important implications for disease prevention strategies, as it suggests not only that disease will propagate more quickly and efficiently on a scale-free network of social contacts than it will on a random network of social contacts, but also that prevention strategies should target those individuals with the most social contacts (those nodes of highest degree in the network), since scale-free networks are most vulnerable to attack of their highest degree nodes [Albert, Jeong and Barabasi, 2000; Dezso and Barabasi, 2002; Gallos, Cohen, Argyrakis, Bunde and Havlin, 2005].  Liljeros *et al.* [2001] have demonstrated that these ideas could be of particular importance for the prevention of sexually-transmitted diseases, as they have shown that the network of human sexual contacts is scale-free; thus, gearing safe-sex education programs toward those individuals with the largest number of sexual partners is likely to have the most dramatic effect on disease prevention.

In cellular networks there is increasing evidence that correct dynamics and function are rooted in the topology of the regulatory networks. For example, Albert and Othmer [2003] have shown that a detailed reconstruction of the regulatory interactions among a group of genes, paired with a coarse-grained description of these genes' functional states, leads to a remarkably successful description of both normal (wild type) and mutant biological behavior. The vast majority of recent publications aims both to explore dynamic processes on networks, as well as to extend static graph-theoretical measures to the description of these dynamics.  The work of Ma'ayan, *et al.*, on cellular networks [2005]**,** and the research of Eckmann [2004] in constructing temporally-sensitive e-mail networks serve as excellent illustrations of initial attempts both to extract dynamic properties from network topologies, as well as to quantify changes in topology due to dynamic processes.  As we continue to explore complexity in fields as diverse as sociology, physics, and biology, network dynamics—which will be addressed in detail in the following two tutorials-- promises to offer the next exciting step in graph-theoretical modeling and applications.

The authors gratefully acknowledge support from the CEMBA IGERT center at the Pennsylvania State University, the National Science Foundation (Grant DMI-0537992) and a Sloan Research Fellowship (to R.A.).